\documentclass[structabstract]{aa}  
\usepackage{graphicx}
\usepackage{txfonts}
\usepackage{natbib}
\usepackage{longtable}
\begin{document}

\title{HIP 10725: The First Solar Twin/Analogue Field Blue Straggler\thanks{Based on observations obtained at the
European Southern Observatory (ESO) Very Large Telescope (VLT) at Paranal Observatory, Chile (observing programs 083.D-0871,
082.C-0446, 093.D-0807), and complemented with observations taken at the Observat\'orio Pico dos Dias (OPD), Brazil (program OP2014A-011).}
}
\titlerunning{HIP 10725: Evidence For a Solar Analogue Field Blue Straggler}

\newcommand{\teff}{$T_{\rm eff}$ }
\newcommand{\tsin}{$T_{\rm eff}$}
\newcommand{\tef}{T_\mathrm{eff}}
\newcommand{\logg}{\log g}
\newcommand{\feh}{\mathrm{[Fe/H]}}

\author{
Lucas Schirbel\inst{1} \and
Jorge Mel\'endez\inst{1} \and
Amanda I. Karakas\inst{2} \and
Iv\'an Ram{\'{\i}}rez\inst{3} \and
Matthieu Castro\inst{4} \and
Marcos A. Faria\inst{5} \and
Maria Lugaro\inst{6} \and
Martin Asplund\inst{2} \and
Marcelo Tucci Maia\inst{1} \and
David Yong\inst{2} \and
Louise Howes\inst{2} \and
J. D. do Nascimento Jr.\inst{4,7}
}


\institute{
Universidade de S\~ao Paulo, Departamento de Astronomia do IAG/USP, Rua do Mat\~ao 1226, Cidade Universit\'aria, 
05508-900 S\~ao Paulo, SP, Brazil. e-mail:  lucas.schirbel@usp.br
\and
The Australian National University, Research School of Astronomy and Astrophysics, Cotter Road, Weston, ACT 2611, Australia
\and
University of Texas at Austin, McDonald Observatory and Department of Astronomy, USA     
\and
Universidade Federal do Rio Grande do Norte, Departamento de F{\'{\i}}sica Te\'orica e Experimental, 59072-970 Natal, RN, Brazil
\and
Universidade Federal de Itajub\'a, DFQ - Instituto de Ci\^encias Exatas, Itajub\'a, MG, Brazil
\and
Konkoly Observatory, Research Centre for Astronomy and Earth Sciences, Hungarian Academy of Sciences, Konkoly Thege Mikl\'os \'ut 15-17, H-1121 Budapest, Hungary
\and
Harvard-Smithsonian Center for Astrophysics, Cambridge, Massachusetts 02138, USA
}

\date{Received ...; accepted ...}

 
  \abstract
   {Blue stragglers are easy to identify in globular clusters, but are much harder 
   to identify in the field. Here we present the serendipitous discovery of one field blue straggler, HIP 10725,  
   that closely matches the Sun in mass and age, but with a metallicity slightly lower than the Sun's. 
   }
   {To characterise the solar twin/analogue HIP 10725 in order to assess whether 
   this star is a blue straggler.
}
   {We employ high resolution (R $\sim 10^5$) high S/N (330) VLT/UVES spectra
   to perform a differential abundance analysis of the solar analogue HIP 10725.
   Radial velocities obtained by other instruments were also used to check 
  for binarity. We also study its chromospheric activity, age and rotational velocity.}
   {We find that HIP 10725 is severely depleted in beryllium ([Be/H] $\leq$ $-1.2$ dex) for its stellar parameters and age. 
   The abundances relative to solar 
   of the elements with Z $\leq$ 30 show a correlation
   with condensation temperature and the neutron capture elements produced by the s-process are greatly enhanced,
   while the r-process elements seem normal.
   We found its projected rotational velocity (\textit{v} sin \textit{i} = 3.3 $\pm0.1$ \,km/s) 
   to be significantly larger than solar, and incompatible with its isochrone-derived age. Radial velocity monitoring shows that the star has a binary companion. 
}
   {Based on the high s-process element enhancements and low beryllium abundance,
we suggest that HIP 10725 has been polluted by mass-transfer from an AGB star,
probably with initial mass of about 2 M$_\odot$.
The radial velocity variations suggest the presence of an unseen binary companion,
probably the remnant of a former AGB star. Isochrones predict
a solar-age star, but this is in disagreement with the high projected rotational velocity 
and high chromospheric activity. We conclude that HIP 10725 is a field blue straggler, 
rejuvenated by the mass transfer process of its former AGB companion.  
}


\maketitle

%

\section{Introduction}

Blue stragglers are traditionally recognised as main-sequence stars, which are significantly bluer than 
the main-sequence turn-off of the population to which they belong \citep{rya01}; 
this phenomenon has been also observed in redder stars \citep[e.g.][]{sil00}. 
Although first identified observationally in clusters \citep{san53},
they are also present in the field \citep[e.g.,][]{car81}.
Field blue stragglers are harder to identify than their cluster counterparts, 
since it is a lot more difficult to reliably establish which stars share a common origin in this case 
(since field stars which initially belonged to a cluster have already dispersed out around the galaxy). 
However, there are other clues left behind by the blue straggler formation processes which can aid in their identification, especially in the case of cooler stars.

Chromospheric activity in blue stragglers tends to be incompatible with their isochrone-derived ages; 
they present significant rotational velocities, thus appearing to be younger \citep{fuh99}. 
Severe Li depletion is also observed, a sign of old age \citep{mon13,mel14a} 
which is also in contradiction with the young age scenario.
When the phenomenon occurs through mass transfer events (i.e., a mass exchange between a now "dead" star during its AGB phase, which pollutes its companion's photosphere, in the process also transferring angular momentum \citep{mcc64}), enhancement of s-process elements is predicted for a sufficiently massive AGB donor, as well as abundance anomalies of light elements \citep{des07}.
In this case, one also expects to find radial velocity variations due to what is now a white dwarf companion. Although these properties could be found individually in normal stars, their presence alone is not an indication of the blue straggler status; their combination, however, strongly suggests so \citep{roc02}.

In the case of solar analogs the argument can be further strengthened by analysing Be abundances.
Large beryllium depletion is not observed in solar twins, where its abundance is relatively constant 
with age \citep{tuc15}\footnote{Only 4 stars in the sample of 118 solar analogs studied by \cite{tak11} are 
extremely depleted in Be. The origin of this depletion is under investigation \citep{via12,des15}.}, 
unlike lithium, which is continuously being depleted as the star 
evolves \citep{bau10,mel10,mel14a,mon13} by extra-mixing below the convective zone. 
This suggests that when Be depletion is present in these stars, other mechanisms must be responsible for it.
The blue straggler scenario is a possible explanation. A small Be depletion would be 
due to the mass transferred by the AGB companion, as it would already 
be devoid of this element. However, a larger Be depletion could be produced 
either by enhanced internal mixing due to the transfer of angular momentum, or to a 
complete destruction of Be as a result of a merger.

In this work, we present evidence for the blue straggler status of the solar twin/analogue 
HIP 10725, being the field blue straggler whose 
stellar parameters (\tsin, log $g$ and metallicity) are the closest to solar.
This star could be classified as a solar twin if the volatile elements (C, N, O) are used 
to characterize its metallicity, or as a solar analog if the depleted refractory elements
(such as iron) are used instead.

\section{Observations}

In our quest to identify new solar twins, we chose eight candidates to be studied at greater detail
(HIP 1536, HIP 3238, HIP 10725, HIP 11514, HIP 106288, HIP 109381, HIP 114328 and HIP 117499),
based on their colours and Hipparcos parallaxes (i.e., their position on the Colour-Magnitude Diagram).
One of those stars (HIP 10725) showed an unusually low Be abundance, 
which sparked further analyses.  
  
The observations for this sample are described in \cite{mel14a}. Briefly, we observed
with UVES@VLT in dichroic mode, with the 346 nm setting (306-387 nm) in the blue arm and the 
580 nm setting (480 - 682 nm) in the red arm. 
Most employed spectral lines are located in the red arm, where we achieved 
R = 110\,000 (0.3 arcsec slit) and S/N of about 330 per pixel at 600 nm.
In the UV we used a slit of 0.6 arcsec, resulting in R = 65\,000.
A reference solar spectrum was obtained using the asteroid Juno  with identical setup. 
The echelle orders were extracted and wavelength calibrated using IRAF, and further processing 
was performed with IDL. 

We also monitored radial velocities for this star to detect possible variations, and were able to acquire four measurements at different epochs.
The first one was the value published by \cite{jen11} based on a FEROS/ESO spectrum taken in October 2008. Our UVES/VLT data described above was obtained in August 2009. Furthermore we obtained two more recent measurements, one taken in July 2014 with UVES/VLT, and another one in August 2014 using the Coud\'e spectrograph at the Observat\'orio Pico dos Dias (OPD) in Brazil. 
The radial velocities are reported in Table 1 and are discussed in Sect. 4.

\begin{figure}
\resizebox{\hsize}{!}{\includegraphics{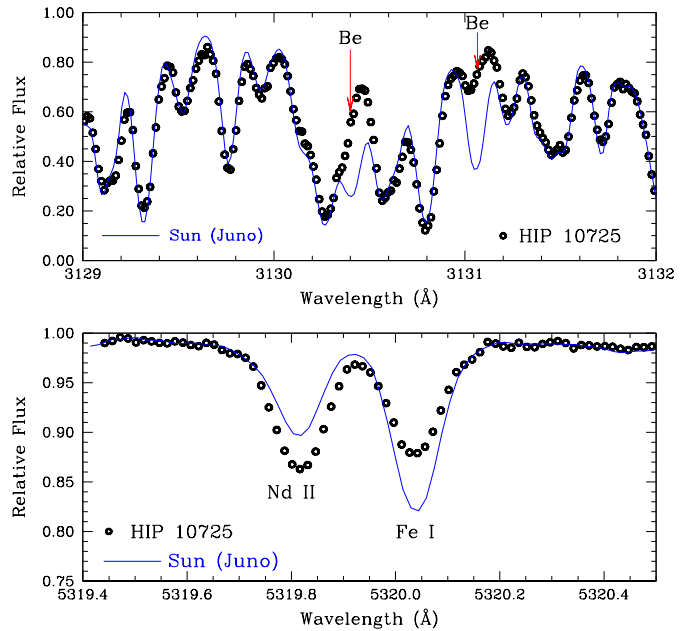}}
\caption{
Comparison of the spectra of HIP 10725 (circles) and the Sun (solid line).
In the top panel, we focus on two BeII features in the UV region around 3030 \AA.
HIP 10725 shows extreme Be depletion compared to the Sun, which is not expected for solar analogues \citep{tuc15}.
The bottom panel shows an enhancement of the neutron capture element Nd in HIP 10725 as well as a substantially lower iron content relative to the Sun.
}
\label{spectra}
\end{figure}

Fig.~\ref{spectra} shows part of the reduced spectra of both the Sun and HIP 10725 around the beryllium feature (\textit{top panel}) and around 5320 \AA\  (\textit{bottom panel}). Given the similarity in stellar parameters, the Be depletion in HIP 10725 is clearly visible, as well as the enhanced Nd abundance and lower iron content.
 
\section{Abundance analysis}

The same differential approach as in our previous papers \citep[e.g.][]{mel14a,ram14} was employed to obtain 
stellar parameters and chemical abundances, i.e., we followed a strictly differential
line-by-line analysis, using ATLAS9 model atmospheres \citep{cas04} and the 
2002 version of the chemical abundance analysis program MOOG \citep{sne73}. 
The equivalent width (EW) measurements were performed manually using IRAF's splot task.
The line list used is presented in \cite{mel14b}.

The differential spectroscopic equilibrium of HIP 10725 relative to the Sun 
results in stellar parameters of \teff = $5777 \pm 16$\,K ($\Delta$ \teff =$ +5 \pm 16$\,K), 
$\log g = 4.45 \pm 0.05$ dex
($\Delta \log g = +0.01 \pm 0.05$ dex), 
[Fe/H] = $-0.17 \pm 0.01$ dex, and a microturbulent velocity of 0.97 km s$^{-1}$ ($+0.07 \pm 0.04$\,km s$^{-1}$ higher than solar).
The uncertainties in the stellar parameters are based on the observational
uncertainties. As the stellar parameters are interdependent of each other, we also took
into account in the error budget this degeneracy.

With these stellar parameters, we computed differential abundances
using the measured EWs, except for Li and Be which were analysed by spectral synthesis following the procedure described 
in Monroe et al. (2013). For Be the
2014 version of MOOG and the line list of Tucci Maia et al. (2015) were employed. 
Hyperfine structure was taken into account
for V, Mn, Co, Y, Ba, La, Pr, Eu and Yb.
The differential abundances are provided in Table \ref{abund} along with the uncertainties stemming 
from observations (standard error in abundances) and stellar parameters. The total error was 
obtained by adding in quadrature the observational and parameter uncertainties.

Notice from the abundance pattern given in Table \ref{abund}, that the 
average of the abundant volatile elements C, N, O, are within 0.1 dex of that of the Sun, 
but that the iron abundance is lower. This has implications regarding the classification 
of a star as  either a solar twin or a solar analog.
Solar twins are classified as stars having \teff within 100 K of the solar effective temperature,
and with log $g$ and [Fe/H] within 0.1 dex of the Sun's \citep{ram09}. [Fe/H] is used as a proxy
of metallicity because it is easier to measure than other more abundant elements such as oxygen
and carbon.
The definition of solar twin could depend on the element chosen to characterise metallicity.
If we consider the average of the abundant volatile elements (C, N, O), then HIP 10725 would be classified
as a solar twin, as its metallicity would be within 0.1 dex of that of the Sun, 
but if the metallicity is based in the depleted refractories (e.g., Fe),
then HIP 10725 would be a solar analog. In any case, HIP 10725 is a star
that closely resembles the Sun.

We estimated the projected rotational velocity $v_{\rm rot} \sin i$ for this star using the 
same procedures and line list as in \cite{tuc15}.
First, we obtained the macroturbulence velocity $V_{\rm macro}$ = 3.6 km s$^{-1}$ 
using the relation given in the aforementioned work. With the macroturbulence fixed, we determined $v_{\rm rot} \sin i = 3.3 \pm 0.1$\,km s$^{-1}$ by spectral synthesis of six lines (5 Fe I lines and 1 Ni I line). Thus the
rotational velocity of HIP 10725 is significantly larger than the solar value (1.9\,kms$^{-1}$).

\begin{figure}
\resizebox{\hsize}{!}{\includegraphics{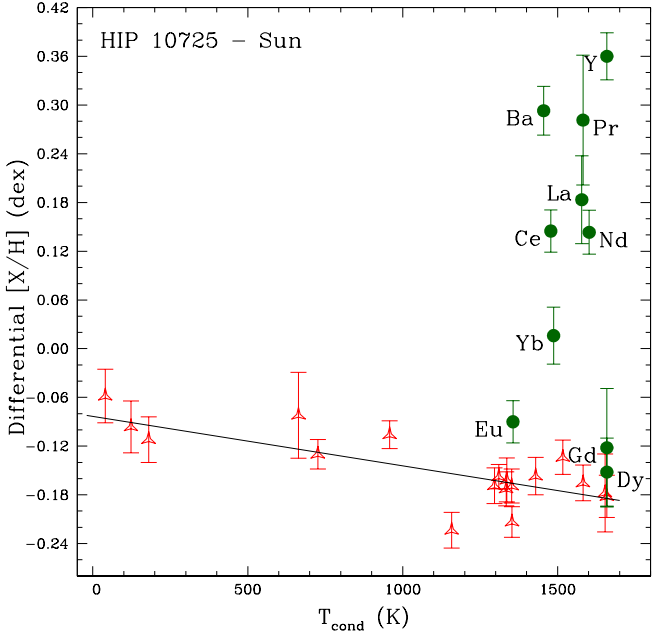}}
\caption{
Differential abundances of HIP 10725 relative to the Sun. The triangles represent the lighter elements ($Z \leq 30$) 
while the filled circles represent neutron-capture elements with $Z > 30$. The solid line is a fit of [X/H] versus $T_{\rm cond}$, 
taking into account only the lighter elements. The fit yields [X/H]$_{(Z \leq 30)}$ = $-0.0834 - 6.10$ $\times 10^{-5}$ $T_{\rm cond}$. 
}
\label{tcond}
\end{figure}

\section{Discussion}

The abundance pattern of HIP 10725 is peculiar in the sense that we observe an excess of s-process elements, 
no enhancement of the r-process elements, and also a trend with condensation temperature ($T_{\rm cond}$), as can be seen in Fig.~\ref{tcond}. 
Usually, this trend is positive in stars closely resembling the Sun, i.e. there is an overabundance of 
refractory elements relative to the Sun \citep{mel09,ram09,ram10}. In this star, 
however, we observe the opposite. In our previous works on solar twins, we speculated that 
the depletion of refractory elements was associated with the formation of rocky planets, 
but in the context of the present work, as suggested by \cite{des07}, the correlation is more 
likely associated with dust formation in the outskirts of the former AGB companion \citep{van03,wat92}:
some of the refractory elements formed dust that was removed from the gas accreted onto HIP 10725. 

Figure 2 shows that the elements that depart the most from the overall $T_{\rm cond}$ trend are those that 
are known to have important contributions from the s-process to their solar system abundances (e.g., Ba, La, Y), 
whereas those with the lowest s-process contribution to those abundances (e.g., Eu, Dy, Gd) 
fall very close to the $T_{\rm cond}$ trend (see for example \cite{sim04}).

\begin{figure}
\resizebox{\hsize}{!}{\includegraphics{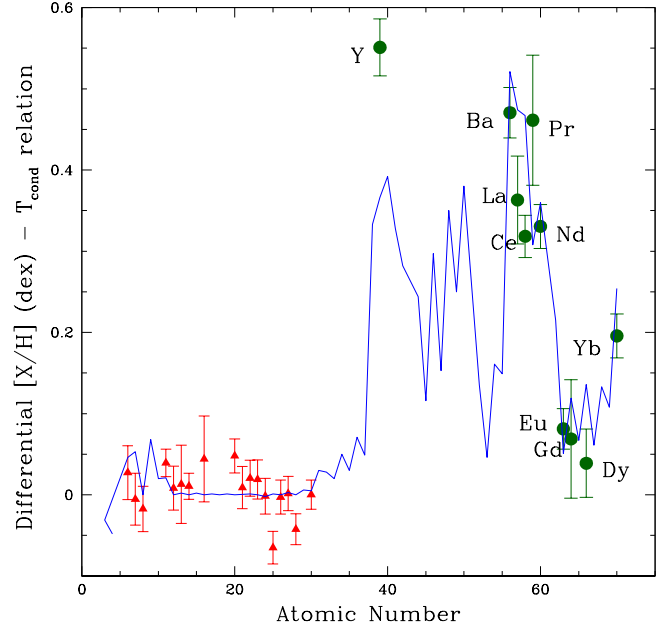}}
\caption{
Differential abundances of HIP 10725 relative to the Sun after the trend with condensation 
temperature has been subtracted. The heavier neutron-capture elements (filled circles) exhibit abundance enhancements,
which are fairly well reproduced by mass transfer from an AGB star with initial mass of 2 M$_\odot$
(solid line), except for yttrium, that is underproduced by the models.
}
\label{trend}
\end{figure}

In order to study this peculiar abundance pattern, we follow the procedure outlined in \citet{mel14b}: 
we fit [X/H] versus condensation temperature \citep{lod03} for the lighter elements (Z $\leq$ 30), 
and subtract this from all abundances. 
The fit can be seen in Fig.~\ref{tcond} and yields $[X/H](Z \leq 30) = -0.0834 - 6.10 \times 10^{-5} T_{\rm cond}$, 
with an element-to-element scatter of 0.03 dex. Fig.~\ref{trend} shows the new abundance ratios after subtraction.
Notice that while the s-process elements like Ba, Nd and Y are enhanced, the r-process elements Eu, Gd and Dy are not,
suggesting that indeed the pollution is due to a former AGB companion.

We modeled the pollution by the former AGB star using solar-metallicity models by \cite{kar14}.
We calculated the $s$-process nucleosynthesis using the same method outlined in
\cite{kar10} and \cite{lug12}.
We mixed the AGB ejecta into the solar analog's shallow surface convection zone, which 
we assumed 0.023M$_\odot$ (2.3\% of the stellar mass), similar to the solar convection zone. 
An AGB stellar model with initial mass of 2 M$_\odot$ provides a reasonable fit to the observed abundance pattern,
as seen in Fig.~\ref{trend}. The amount of accreted AGB material is 0.2\% of the mass lost 
from the 2 M$_\odot$ AGB star over its lifetime, 
with most of the material lost at the tip of the AGB during the last couple of thermal pulses. 
This choice provides a fair fit
to Ba, Nd, Eu, Gd, Yb, and Dy. The predicted Y abundance is lower than observed,
but this discrepancy could be related to the formation of 
$^{13}$C pockets in AGB stars or to the activation of the $^{22}$Ne neutron source 
in the thermal pulses (see review by \cite{kl14}).

\begin{figure}
\resizebox{\hsize}{!}{\includegraphics{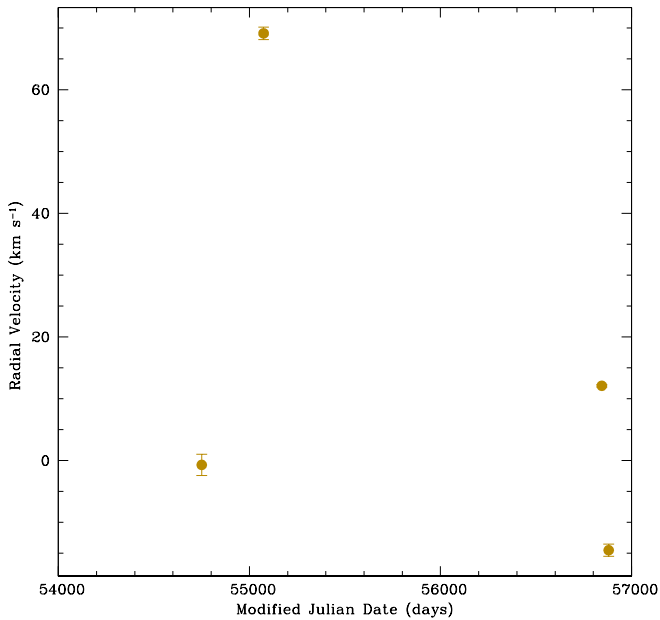}}
\caption{Plot showing radial velocity data obtained for HIP 10725 over the course of several years, evidence for the presence of the unseen white dwarf companion.  
}
\label{radial}
\end{figure}

As shown above, mass transfer from an AGB star can account for the enhanced s-process pattern observed, 
as expected for blue stragglers formed by this process \citep{des07}. 
The donor star would subsequently have evolved into an unseen white dwarf companion. 
Such a companion star should be detectable through radial velocity monitoring, 
which is indeed the case. As seen in Fig.~\ref{radial}, we found radial velocity variations 
of up to $\sim 70$\,km s$^{-1}$, good evidence of the binarity of the system, 
which is common in the case of field blue stragglers \citep{pre00,car05}. 
The radial velocities and observation dates are presented in Table 1.

In addition, we also determined Be abundances for HIP 10725 using spectral synthesis, 
and found a value of A(Be) $\leq$ 0.2 dex, or [Be/H] $\leq$ $-1.2$ dex, for A(Be)$_{\odot}$ = 1.38 dex \citep{asp09},
meaning that the beryllium abundance in this star is at least 15 times lower than in the Sun. 
This is also compatible with the AGB mass transfer scenario. 
In the AGB scenario we propose, Be depletion would be small (0.05 dex) due to the Be-depleted 
material transfered by the AGB star, and large (1.15 dex) due to the transfer of angular momentum 
from the AGB star, which would cause the solar twin to spin up, as suggested by its higher rotational velocity.

In order to study the depletion of Be we use the Toulouse-Geneva Evolution Code \citep[TGEC,][]{hui08}
including standard physics and also extra mixing beyond that predicted by standard models.
We take into account diffusion (including gravitational settling)
and rotation-induced mixing, as described in our previous works \citep{nas09,cas11,tuc15}.
We modeled the Be depletion using as parameters the  diffusion coefficient at 
the base of the convective zone $D_{bcz}$, and the half-height width of the tachocline $\Delta$,
which are both free parameters of stellar evolution models.
 
The mass transfer from the AGB star
into the surface of HIP 10725 should have enhanced the angular 
momentum of the convective envelope, increasing the stellar rotation.
Convection transfers this increase until the base of the convective zone, where exists a transition 
zone between the differential rotation of the convective zone and the interior that rotates as a solid body. 
This transition zone, called tachocline \citep{spi92,ric96,bru98}, 
induces a mixing at the base of the convective zone which can be modelled by an 
exponential effective diffusion coefficient $D_{bcz}$. 
In order to reproduce the observed Be depletion, 
we multiplied $D_{bcz}$ by 1.7 and increased the size of the tachocline $\Delta$ by a factor of 2.3. Doing so 
we can match the observed destruction of Be. The above changes in $D_{bcz}$ and $\Delta$, result
in a surface velocity of 3.7 km s$^{-1}$, close to the observed  $v_{\rm rot} \sin i$ = 3.3 km s$^{-1}$.

Enhanced chromospheric activity ($R_{HK} = $\ $-$4.51, \cite{jen11}) and high rotational velocity 
(both suggesting a young star) are incompatible with the isochrone-derived age obtained for HIP 10725 \citep{kim02,dem04}, 
using probability distribution functions as in \cite{mel12}. Our determined value of $v_{\rm rot} \sin i$ = 3.3 km s$^{-1}$, 
suggests an age of roughly 1 Gyr, while the Yonsei-Yale isochrones result in an age of 5.2$_{-2.1}^{+1.9}$ Gyr. 
Our upper limit for the Li abundance (A(Li) $\leq$ 0.9 dex) is also about 20 times smaller than the
Li abundance expected for an age of 1 Gyr \citep{tuc15}, showing that the star is not young. 
Thus, the high projected rotational velocity is most likely the consequence 
of angular momentum transfer during the accretion stage from the former AGB star \citep{fuh99},
that ``rejuvenated'' HIP 10725, enhancing both rotation and chromospheric activity.

\section{Conclusions}
We present an observational study demonstrating that the solar analog HIP 10725 
has been polluted by s-process material. 
Its high s-process element abundance is in good agreement with mass transfer from a 2-M$_\odot$ AGB star.
A white dwarf should remain from the now defunct AGB star, and we have been able to detect
the presence of a companion, through radial velocity variations. While the 
isochrone-derived age 
shows that its age is similar to solar, its chromospheric activity 
and a higher than usual rotational velocity would instead suggest that the star is young. 
This "rejuvenation" process is a consequence of mass and angular momentum transfer via wind accretion or 
Roche lobe overflow during the accretion period. 
A large beryllium depletion compared with solar-type stars of similar mass and age is present as well, 
providing further evidence for transfer of angular momentum having taken place.

Thus, we identify HIP 10725 as a field blue straggler. Future acquisition of more radial velocity measurements will help better constrain orbital parameters, 
and determine the mass of the probable white dwarf companion.

Finally, our work may be relevant to explain the large depletion in beryllium observed 
in four of the 118 solar analogs observed by \cite{tak11},
as some of these stars could also be field blue stragglers.
Indeed, in a recent work, \cite{des15} shows that those four Be-depleted solar analogs
have binary companions, and for two of them the companion is a white dwarf, showing
thus a connection with the blue straggler phenomenon.

\begin{acknowledgements}

J.M. would like to acknowledge support from
FAPESP (2012/24392-2) and CNPq ({\em Bolsa de Produtividade}).
L.S. acknowledges support from FAPESP (2013/25008-4) and CNPq.
M.A and D.Y acknowledge support from the Australian Research Council (grants FL110100012 and DP120100991).
M.L. is a Momentum project leader of the Hungarian Academy of Sciences.
M.C. acknowledges the support grant 311706/2014-2 from CNPq.
AK was supported through an Australian Research Council Future Fellowship (FT110100475).
\end{acknowledgements}

\clearpage

\begin{table}
\caption{Radial velocity measurements}
\centering 
\renewcommand{\footnoterule}{}
\begin{tabular}{lrrrrrrrrrr}
\hline
\hline

{M.J.D.}\tablefootmark{a} & Radial Velocity & Error & Date \\
(days)  & (km s$^{-1}$) & (km s$^{-1}$) & (yyyy-mm-dd) \\
\hline

54750.270436 & $-00.70$ &  1.7 & 2008-10-11 \\
55074.321678 & $ 69.10$ &  1.0 & 2009-08-31 \\
56844.436451 & $ 12.10$ &  0.2 & 2014-07-06 \\
56880.270833 & $-14.50$ &  1.0 & 2014-08-11 \\
\hline

\end{tabular}
\tablefoot{\tablefoottext{a}{Modified Julian Date}}
\end{table}

\begin{table}
\caption{Differential abundances\tablefootmark{*} of HIP 10725 relative to the Sun and errors in abundances $\Delta$A
due to observational and systematic (stellar parameters) uncertainties}
\label{abund}
\centering 
\renewcommand{\footnoterule}{}  
\begin{tabular}{lrrrrrrrrrr} 
\hline
\hline
{}& [X/H]   & $\Delta$A/$\Delta \tef$ & $\Delta$A/$\Delta$log $g$ & $\Delta$A/$\Delta v_t$ & $\Delta$A/$\Delta$[Fe/H] & error & error & error \\
\hline 
{Element}& LTE   & $\Delta \tef$ & $\Delta$log $g$ & $\Delta v_t$ & $\Delta$[Fe/H] & param\tablefootmark{a} & obs\tablefootmark{b} & total\tablefootmark{c} \\
{}       &       &   +16K           &  +0.05 dex      & +0.04 km s$^{-1}$  & +0.01 dex   &  &  &  \\
{}       & (dex) & (dex) & (dex)         & (dex)           & (dex)       & (dex)        & (dex) & (dex) \\
\hline
C  & $-0.058$ & $-0.008$ & $ 0.011$ & $ 0.000$ & $ 0.001$ & $ 0.013$ & $ 0.030$ & $ 0.033$ \\
N  & $-0.096$ & $ 0.022$ & $ 0.007$ & $ 0.008$ & $-0.007$ & $ 0.025$ & $ 0.020$ & $ 0.032$ \\
O  & $-0.112$ & $ 0.021$ & $ 0.006$ & $ 0.001$ & $-0.007$ & $ 0.023$ & $ 0.017$ & $ 0.028$ \\
Na & $-0.106$ & $ 0.009$ & $ 0.000$ & $-0.001$ & $ 0.000$ & $ 0.009$ & $ 0.015$ & $ 0.017$ \\
Mg & $-0.162$ & $ 0.008$ & $-0.004$ & $-0.002$ & $-0.001$ & $ 0.009$ & $ 0.026$ & $ 0.027$ \\
Al & $-0.178$ & $ 0.008$ & $ 0.001$ & $-0.001$ & $ 0.000$ & $ 0.008$ & $ 0.048$ & $ 0.048$ \\
Si & $-0.158$ & $ 0.004$ & $ 0.004$ & $-0.001$ & $-0.001$ & $ 0.005$ & $ 0.015$ & $ 0.015$ \\
S  & $-0.082$ & $-0.007$ & $ 0.009$ & $-0.001$ & $ 0.000$ & $ 0.011$ & $ 0.052$ & $ 0.053$ \\
Ca & $-0.134$ & $ 0.011$ & $-0.007$ & $-0.007$ & $-0.001$ & $ 0.015$ & $ 0.015$ & $ 0.021$ \\
Sc & $-0.182$ & $ 0.002$ & $ 0.018$ & $-0.005$ & $-0.003$ & $ 0.019$ & $ 0.018$ & $ 0.026$ \\
Ti & $-0.165$ & $ 0.016$ & $ 0.003$ & $-0.004$ & $ 0.000$ & $ 0.017$ & $ 0.015$ & $ 0.022$ \\
V  & $-0.157$ & $ 0.017$ & $ 0.004$ & $-0.002$ & $-0.001$ & $ 0.017$ & $ 0.015$ & $ 0.023$ \\
Cr & $-0.169$ & $ 0.014$ & $-0.001$ & $-0.008$ & $-0.001$ & $ 0.016$ & $ 0.015$ & $ 0.022$ \\
Mn & $-0.223$ & $ 0.013$ & $-0.001$ & $-0.003$ & $ 0.001$ & $ 0.013$ & $ 0.015$ & $ 0.020$ \\
Fe & $-0.173$ & $ 0.013$ & $-0.001$ & $-0.007$ & $-0.001$ & $ 0.014$ & $ 0.015$ & $ 0.021$ \\
Co & $-0.169$ & $ 0.012$ & $ 0.008$ & $ 0.002$ & $ 0.000$ & $ 0.015$ & $ 0.016$ & $ 0.022$ \\
Ni & $-0.213$ & $ 0.010$ & $ 0.002$ & $-0.005$ & $-0.001$ & $ 0.011$ & $ 0.015$ & $ 0.019$ \\
Zn & $-0.130$ & $ 0.002$ & $ 0.005$ & $-0.008$ & $-0.002$ & $ 0.009$ & $ 0.015$ & $ 0.018$ \\
Y  & $ 0.360$ & $ 0.003$ & $ 0.008$ & $-0.021$ & $-0.022$ & $ 0.031$ & $ 0.015$ & $ 0.035$ \\
Ba & $ 0.293$ & $ 0.007$ & $ 0.000$ & $-0.013$ & $-0.007$ & $ 0.016$ & $ 0.026$ & $ 0.031$ \\
La & $ 0.183$ & $ 0.005$ & $ 0.029$ & $ 0.001$ & $-0.005$ & $ 0.030$ & $ 0.045$ & $ 0.054$ \\
Ce & $ 0.145$ & $ 0.004$ & $ 0.021$ & $-0.002$ & $-0.003$ & $ 0.022$ & $ 0.015$ & $ 0.026$ \\
Pr & $ 0.281$ & $ 0.031$ & $ 0.045$ & $ 0.025$ & $ 0.025$ & $ 0.066$ & $ 0.046$ & $ 0.080$ \\
Nd & $ 0.143$ & $ 0.005$ & $ 0.022$ & $-0.003$ & $-0.003$ & $ 0.023$ & $ 0.015$ & $ 0.027$ \\
Eu & $-0.090$ & $ 0.000$ & $ 0.018$ & $-0.002$ & $-0.002$ & $ 0.018$ & $ 0.017$ & $ 0.025$ \\
Gd & $-0.122$ & $ 0.006$ & $ 0.024$ & $-0.001$ & $-0.003$ & $ 0.025$ & $ 0.069$ & $ 0.073$ \\
Dy & $-0.152$ & $ 0.006$ & $ 0.022$ & $-0.007$ & $-0.004$ & $ 0.024$ & $ 0.035$ & $ 0.042$ \\
Yb & $ 0.016$ & $ 0.003$ & $ 0.021$ & $-0.007$ & $-0.005$ & $ 0.023$ & $ 0.015$ & $ 0.027$ \\
\hline       
\end{tabular}
\tablefoot{Abundances of V, Mn, Co, Y, Ba, La, Pr, Eu and Yb account for HFS.\\
\tablefoottext{*}{A(Li) $\leq$ 0.9 dex, A(Be) $\leq$ 0.2 dex }
\tablefoottext{a}{Adding errors in stellar parameters}
\tablefoottext{b}{Observational errors}
\tablefoottext{c}{Total error (stellar parameters and observational)}
}
\end{table}

\end{document}